\documentclass[fleqn,12pt,twoside]{article}
\usepackage{espcrc1}
\input epsf
\title{SPIN PHYSICS AT JLAB IN THE RESONANCE REGION}

\author{VOLKER D. BURKERT \\
Jefferson Lab, 12000 Jefferson Avenue, Newport News, VA23606}
\begin{document}
\maketitle
\vspace{-4.0cm}
\hspace{12cm}JLAB-PHY-01-19
\vspace{4.cm}
\abstract{I will discuss recent results from Jefferson Lab on the measurement 
of inclusive spin structure functions in the nucleon resonance region 
using polarized ammonia $NH_3$ and polarized $^3He$ targets. Preliminary 
results on the first moment of $g_1(x,Q^2)$ for protons, and the 
generalized Gerasimov-Drell-Hearn integral for neutrons are presented.
In addition, first double polarization data on exclusive electroproduction of  
$\pi^+$ from polarized protons will be discussed.}

\section{INTRODUCTION}
Since the mid 1970's measurements of polarized structure functions in lepton 
nucleon scattering have been a focus of nucleon structure physics at
high energy laboratories. One of the surprising  findings 
was that less than 25\% of the nucleon spin is accounted for by 
the spin of quarks \cite{filipone}. At asymptotic momentum and energy 
transfers the 
Bjorken sum rule \cite{bjorken} relates the proton-neutron difference 
of the first moment $\Gamma_1 = \int{g_1(x)dx}$ to 
the weak axial coupling constant $g_A$: 
$$\Gamma_1^p - \Gamma_1^n = {1\over 6}g_A \eqno(1)$$  

This sum rule has been evolved in perturbative QCD (pQCD) to finite 
$Q^2$, and has been verified experimentally at the 5\% level.
However, the fact, that only a small fraction of the 
nucleon spin can be directly 
attributed to the quark spin is in contrast to quark model expectations, 
and shows 
that we are far from having a realistic
picture of the intrinsic structure of the nucleon. 
Moreover, the nucleon structure has hardly been 
explored in the regime of strong interaction which is the true domain
of QCD. Our understanding of nucleon structure is incomplete 
if the nucleon is not probed and fundamentally described 
at both large and medium distance scales. 
This is the domain where current experiments at JLab have their 
greatest impact.

The Gerasimov-Drell-Hearn (GDH) sum rule\cite{gerasimov,drell} 
relates the differences in the helicity-dependent total photoabsorption 
cross sections to the anomalous magnetic moment $\kappa^2$ of the target 
 
$${M^2 \over 8\pi^2\alpha}\int_{\nu_0}^{\infty}
{{{\sigma_{1/2}(\nu)-\sigma_{3/2}(\nu)}\over \nu}d\nu =
 -{1\over 4}\kappa^2}\eqno(2)$$  
\noindent
where $\nu_0$ is the photon energy at pion threshold, and M is the nucleon mass.
This sum rule has been studied for photon energies up 
to 850 MeV \cite{ahrens}, and is currently being tested up to more 
than 2 GeV at ELSA \cite{anton}.  
A rigorous extension of this sum rule to finite $Q^2$ has recently been introduced \cite{jios}. 
This makes a measurement of the $Q^2$-dependence of (2) very interesting.
It will tell us at what distance scale pQCD corrections and higher twist 
expansions will break down, and physics of confinement will dominate. 
It will also allow us to evaluate where resonances give important, even 
dominant 
contributions to the first moment \cite{burli,buriof}, as well as to the 
higher $x$-moments of 
the spin structure function $g_1(x,Q^2)$~. 
These contributions need to be determined experimentally and calculated 
in QCD. 
The helicity structure of some resonances is changing rapidly 
with the distance scale probed \cite{burk1}. Accurate measurements will allow 
stringent tests of nucleon structure models. 
The well known ``duality'' between the deep inelastic regime and the 
resonance regime observed for the unpolarized structure function $F_1(x,Q^2)$ 
needs to be 
explored for the spin structure function $g_1(x,Q^2)$. This will shed new 
light on this phenomenon which is not well understood.          
It is only with a concerted effort of precise experiments and new 
approaches in theory that we will be able to understand nucleon structure
from the smallest to the largest distances. 
It is one of the goals of experiments at JLab to provide the basis
for such an endeavor.  The first experiments have been completed on polarized
hydrogen ($NH_3$) and deuterium ($ND_3$) targets, and on polarized $^3He$.
%%%%%%%%%%%%%%%%%%%%%%%%%%%%%%%%%%%%%%%%%%%%%%%%%%%
\begin{figure}
\vspace{65mm} 
\centering{\includegraphics{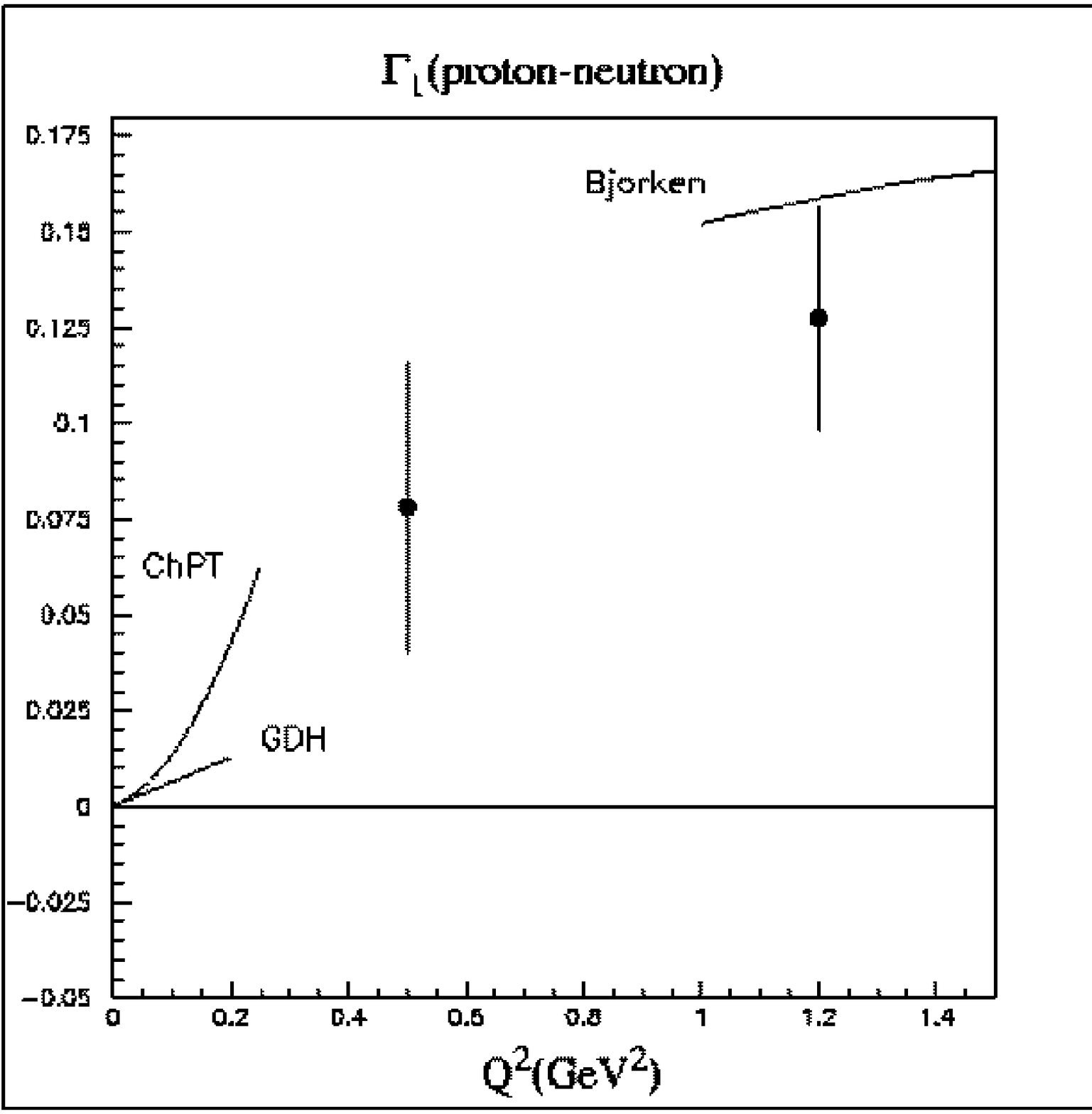}}
\centering{\includegraphics{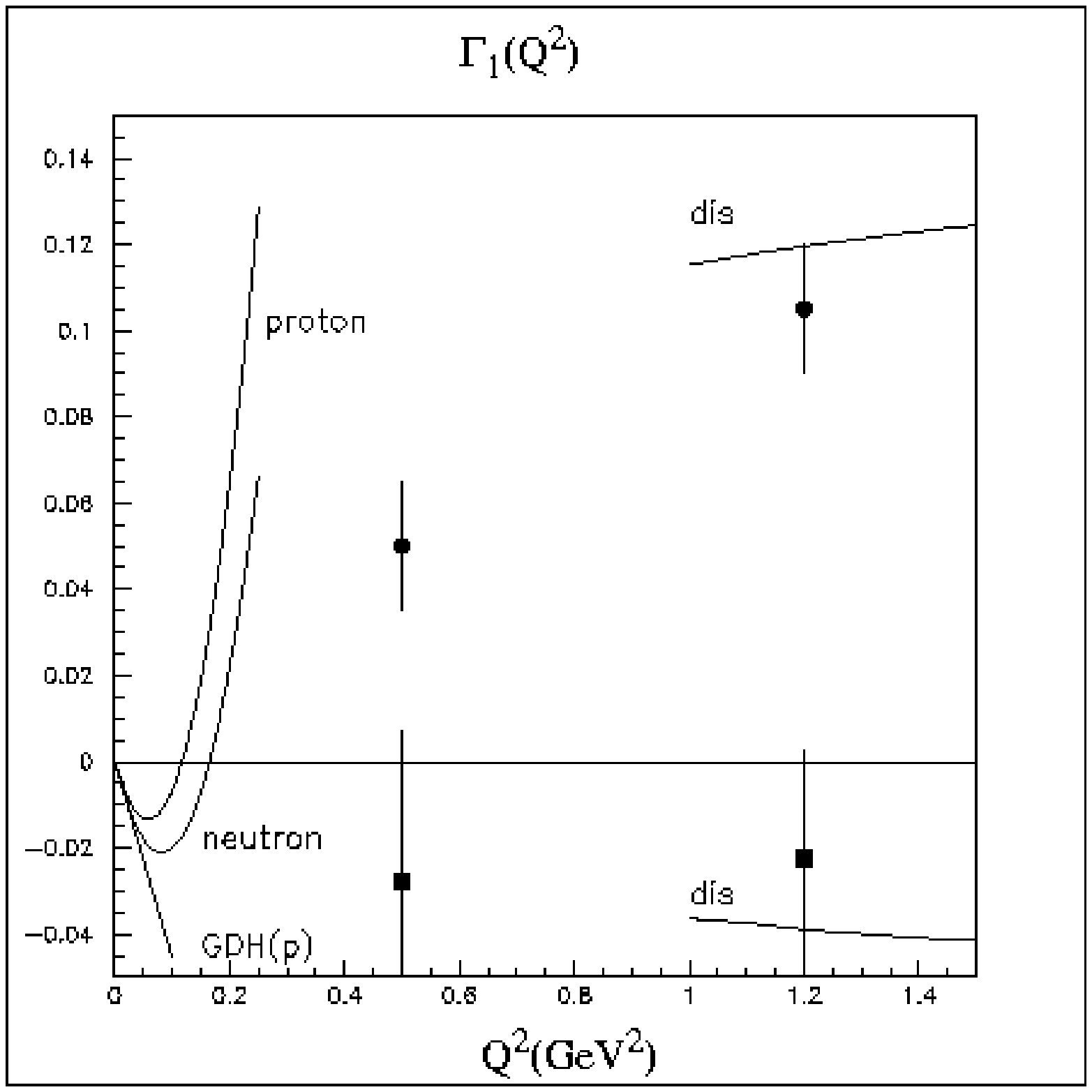}}
\caption{\small First moments of the spin structure 
function $g_1(x,Q^2)$ for the 
proton and neutron (left), and for the proton-neutron 
difference (right). The curves above $Q^2=1$GeV$^2$ are pQCD evolutions of 
the measured $\Gamma_1$ for proton and neutron, and the pQCD evolution for 
the Bjorken sum rule, respectively. The straight lines near $Q^2=0$ are the 
slope given by the GDH sum rule. 
The curves at small $Q^2$ represent the NLO HBChPT results.}.
\label{gamma1}
\end{figure}
%%%%%%%%%%%%%%%%%%%%%%%%%%%%%%%%%%%%%%%%%%%%%%%%%%%%%

\section{EXPECTATIONS FOR  $\Gamma_1(Q^2)$}

The inclusive doubly polarized cross section can be written as:
$${1\over \Gamma_T} {d\sigma \over d\Omega dE^{\prime}} = \sigma_T 
+ \epsilon\sigma_L + P_eP_t[\sqrt{1-\epsilon^2}A_1\sigma_T\cos{\psi} + 
\sqrt{2\epsilon(1+\epsilon)}A_2\sigma_T\sin{\psi}]~\eqno(3)$$
where $A_1$ and $A_2$ are the spin-dependent asymmetries, $\psi$ is the angle between the nucleon polarization vector and the $\vec q$ vector, $\epsilon$ the
polarization parameter of the virtual photon, and $\sigma_T$ and $\sigma_L$ 
are the total absorption cross sections for transverse and longitudinal 
virtual photons. Experiments usually measure the asymmetry 
$$A_{exp} = P_eP_tD {A_1+\eta A_2 \over 1+\epsilon R}\eqno(4)$$
where D is a kinematical factor describing the polarization transfer from the
electron to the photon. 
The asymmetries $A_1$ and $A_2$ are related to the spin structure function $g_1$ 
by  $$g_1(x,Q^2) = {\tau \over 1+\tau}
[A_1 + {1\over \sqrt{\tau}}A_2]F_1(x,Q^2) \eqno(5)$$ 
where $F_1$ is the usual unpolarized structure function. 

The GDH sum rule defines the slope of $\Gamma_1(Q^2=0)$, where we exclude 
the elastic contribution at x=1: 
$$2M^2{d\Gamma_1 \over dQ^2}(Q^2\rightarrow 0) = -{1\over 4}\kappa^2 \eqno(6)$$

%%%%%%%%%%%%%%%%%%% %%%%%%%%%%%%%%%%%%%%%%%%%%%%%%%%
\begin{figure}[t]
\vspace{100mm} 
\centering{\includegraphics{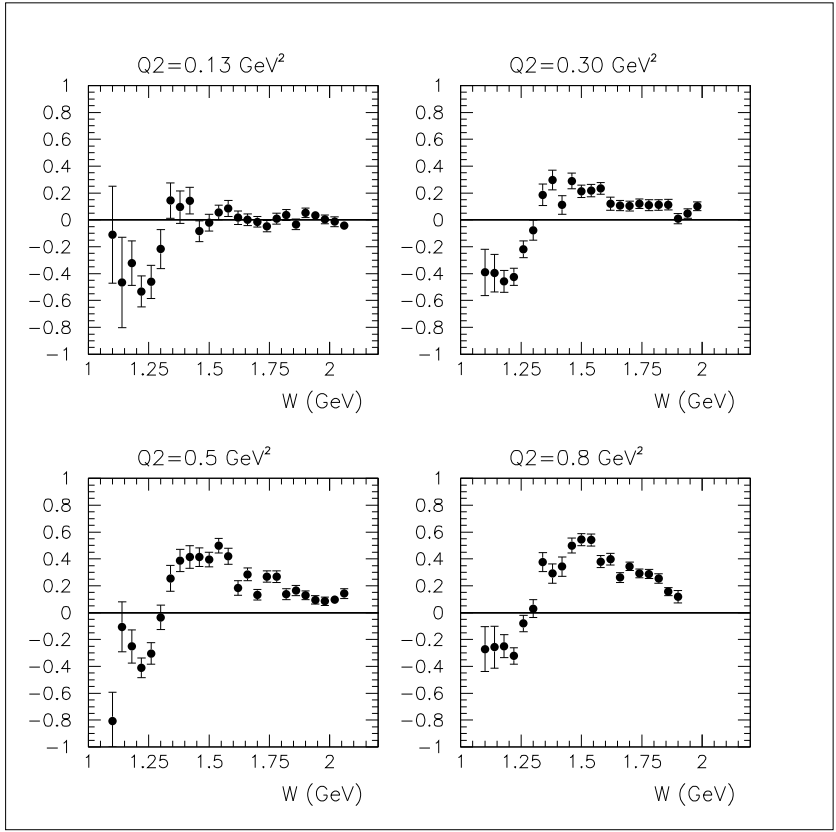}}
\caption{\small Asymmetry $A_1+\eta A_2$ for protons. The 
panels show the results for different $Q^2$ values.}
\label{epasym}
\end{figure}
%%%%%%%%%%%%%%%%%%%%%%%%%%%%%%%%%%%%%%%%%%%%%%%%%%%%%

The GDH and Bjorken sum rules provide constraints at the kinematical 
endpoints $Q^2=0$ and $Q^2 \rightarrow \infty$. 
The evolution of the Bjorken sum rule to finite values of $Q^2$ using 
pQCD and the twist expansion allow to connect experimental values measured at
finite $Q^2$ to the endpoints. 
Heavy Baryon Chiral Perturbation Theory (HBChPT) has been 
proposed as a tool to evolve the GDH sum rule to $Q^2 \neq 0$, possibly to
$Q^2 = 0.1$GeV$^2$, and to use the twist expansion down to $Q^2=0.5$GeV$^2$ 
\cite{ji}. 
The challenge of nucleon structure physics would be to bridge the remaining 
gap from $Q^2$ = 0.1 - 0.5 GeV$^2$. Lattice QCD may play an important role 
in describing resonance contributions to the moments of spin 
structure functions. If successful this would {\sl mark the 
first time that a quantity of nucleon structure physics is described from
small to large distances within fundamental theory}, a worthwhile goal! 
 
Using just the constraints given by the two endpoint sum rules we may already
get a qualitative picture on $\Gamma^p_1(Q^2)$ and $\Gamma^n_1(Q^2)$. 
There is no sum rule for the proton and neutron separately that
has been verified. However, experiments have determined the asymptotic 
limit with sufficient confidence for the proton and the neutron. At large 
$Q^2$, $\Gamma_1$  is expected to approach this limit following the 
pQCD evolution from finite values of $Q^2$. At small $Q^2$, $\Gamma_1$ 
must approach zero with a slope given by the GDH sum rule 
(assuming the sum rule will be verified). 
The current situation is depicted in Figure {\ref{gamma1}}, where also 
the next-to-leading HBChPT evolution at small $Q^2$ and the 
pQCD evolution to order $\alpha_s^3$ at high $Q^2$ are shown.

%%%%%%%%%%%%%%%%%%% %%%%%%%%%%%%%%%%%%%%%%%%%%%%%%%%
\begin{figure}[t]
\vspace{90mm} 
\centering{\includegraphics{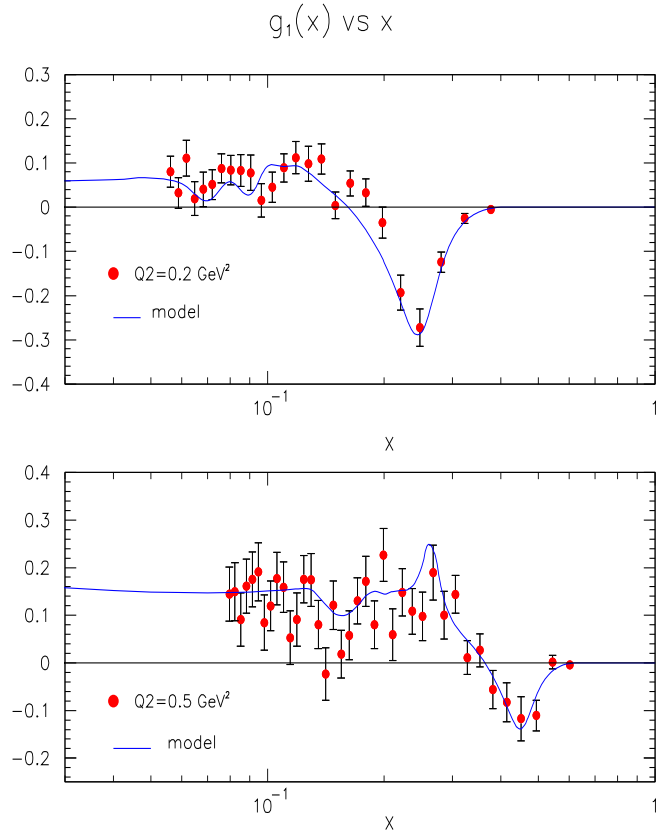}}
\caption{\small Spin structure function $g_1(x,Q^2)$ for the 
proton. The curve labeled ``model'' is used for radiative corrections, and 
to extrapolate to $x=0$ for the evaluation of $\Gamma_1$.}
\label{g1x}
\end{figure}
%%%%%%%%%%%%%%%%%%%%%%%%%%%%%%%%%%%%%%%%%%%%%%%%%%%%%

As the slope at $Q^2=0$ is $< 0$, $\Gamma^p_1$ must change sign at 
some value of $Q^2 < 0.5$GeV$^2$. 
We note that the HBChPT NLO evolution \cite{xji} 
cannot give a good description of the 
trend shown by the existing data, for $Q^2> 0.1$GeV$^2$. 
However, for the proton-neutron difference the situation is quite 
different \cite{burk}; the HBChPT curve describes the 
general trend of the data quite well, and over a significantly 
larger range in $Q^2$ than for proton and neutron separately.

\section{PRELIMINARY RESULTS FOR PROTONS AND NEUTRONS.}

Inclusive double polarization experiments have been carried out on 
polarized hydrogen \cite{bucramin}, 
using $N\vec {H_3}$ as polarized target material. 

In Figure {\ref{epasym}} the asymmetry $A_1+\eta A_2$ is shown 
for various bins at low $Q^2$. 
In the lowest $Q^2$ bin the asymmetry is dominated by 
the excitation of the $\Delta(1232)$. At higher $Q^2$ the asymmetry in the
$\Delta(1232)$ region remains negative, however, at higher W it 
quickly becomes positive and large, reaching peak values of about 
0.6 at $Q^2=0.8$GeV$^2$ and W=1.5GeV. Evaluations of resonance
contributions show that this is largely driven by the 
$S_{11}(1535)$ $A_{1/2}$ amplitude, and by the rapidly changing helicity 
structure of the strong $D_{13}(1520)$ state. The latter resonance is 
known to have a dominant $A_{3/2}$ amplitude at the photon point, but 
is rapidly changing to $A_{1/2}$ dominance for $Q^2 > 0.5$GeV$^2$.

Using a parametrization of the world data on $F_1(x,Q^2)$ and 
$A_2(x,Q^2)$ we can extract $g_1(x,Q^2)$ from (5). Examples of 
$g_1(x,Q^2)$ are shown in Figure {\ref{g1x}}. 
The main feature at low $Q^2$ is due to the negative contribution of the
$\Delta(1232)$ resonance. The graphs also show a model parametrization 
of $g_1(x,Q^2)$ which was used to extrapolate to $x \rightarrow 0$. 
The extrapolation is needed to evaluate the first moment $\Gamma_1(Q^2)$.

\begin{figure}[t]
\vspace{85mm} 
\centering{\includegraphics{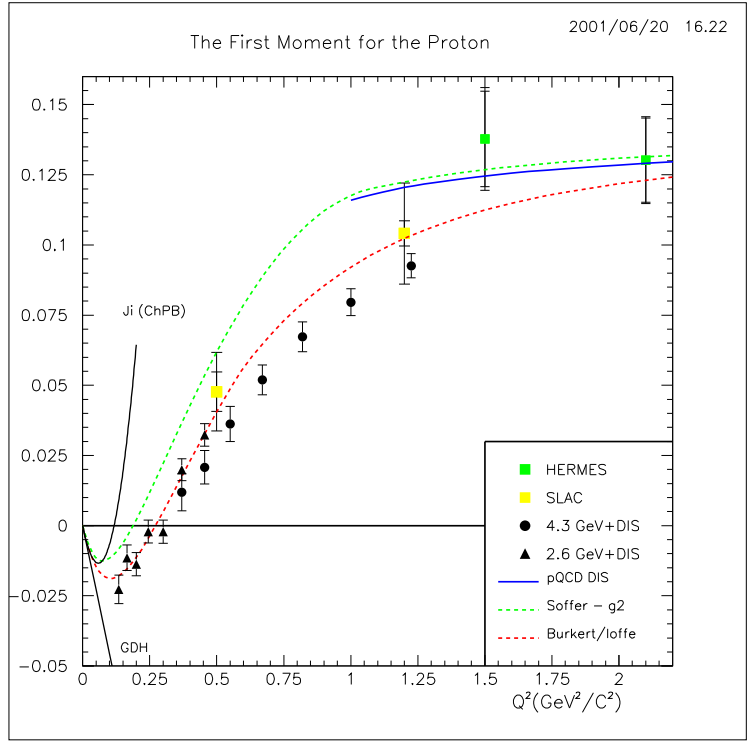}}
\caption{\small First moment $\Gamma_1(Q^2)$ for the proton. The black symbols are preliminary results from CLAS. Data from SLAC and HERMES are shown for 
comparison}
\label{gammafull}
\end{figure}

The results  for $\Gamma^p_1(Q^2)$ are shown in Figure {\ref{gammafull}}. The 
characteristic feature is the strong $Q^2$ dependence for $Q^2 < 1$GeV$^2$, 
with a zero crossing near $Q^2=0.3$ GeV$^2$. This result is preliminary, 
and the final results may change within the systematic uncertainties. However, 
the qualitative features of the data will not change.

%%%%%%%%%%%%%%%%%%% %%%%%%%%%%%%%%%%%%%%%%%%%%%%%%%%
%%%%%%%%%%%%%%%%%%%%%%%%%%%%%%%%%%%%%%%%%%%%%%%%%%%%%

\begin{center}
\begin{figure}
\vspace{90mm} 
\centering{\includegraphics{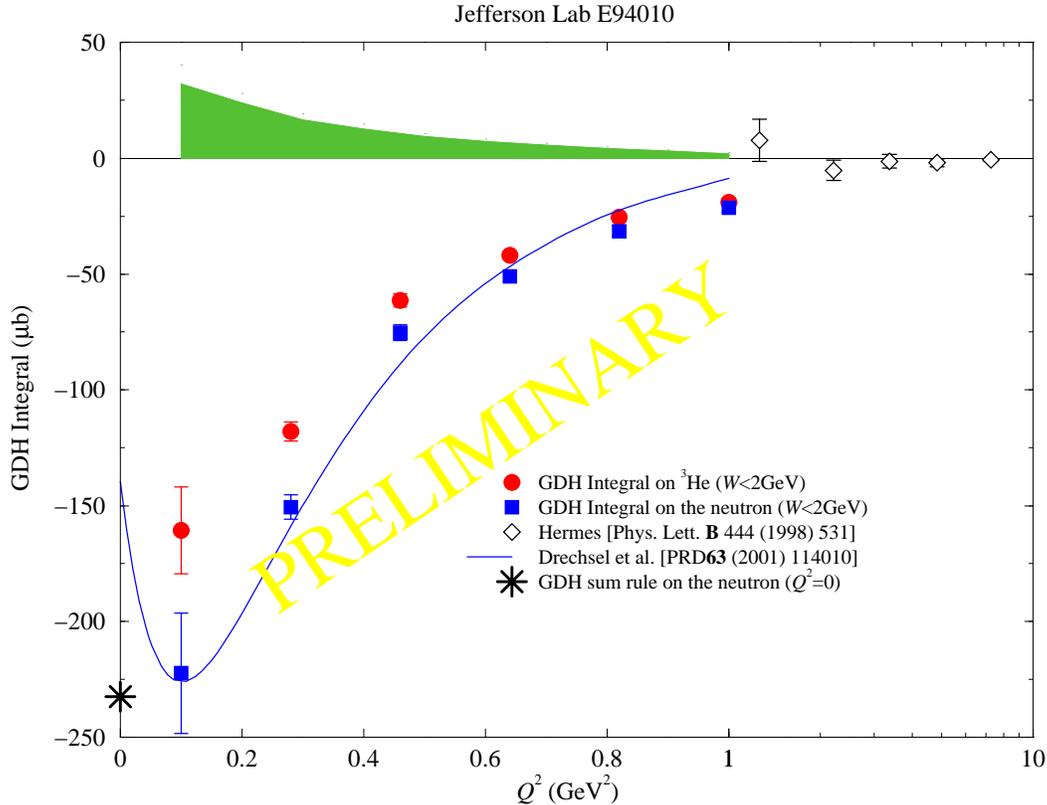}}
\caption{\small Generalized GDH integral for $^3He$ and neutron. The 
dark band indicates the systematic errors. }
\label{gdhneutron}
\end{figure}
\end{center}

Measurements on $ND_3$ have been carried out in CLAS \cite{kuhn}, 
and on  $^3He$ in Hall A \cite{halla1} to measure the 
corresponding integrals for the neutron.
Data were taken with  
the JLab Hall A spectrometers using a polarized $^3He$ target. Since the 
data were taken at fixed scattering angle, $Q^2$ and $\nu$ are correlated.
Cross sections at fixed $Q^2$ are determined by an interpolation between 
measurements at different beam energies. Both longitudinal and transverse 
settings of the target polarization were used. 
Therefore no assumptions about $A_2$ 
are necessary in this case. 
The GDH integral for $^3He$ and for neutrons is shown in Figure 
{\ref{gdhneutron}}. 
The integral is evaluated  
over the region from pion threshold (on a free neutron) 
to W=2 GeV, to cover the resonance region only. 
Corrections for nuclear effects were made based on a prescription by
Ciofi degli Atti \cite{ciofi}.

The GDH integral for the neutron shows a $Q^2$ behavior that seems to 
smoothly approach the GDH sum rule value. However, the data are also in
reasonable agreement with the model curve, which indicates a rapid departure 
away from the sum rule value at very small $Q^2$. Data at smaller $Q^2$ 
and for real photons should clarify the situation.

%%%%%%%%%%%%%%%%%%% %%%%%%%%%%%%%%%%%%%%%%%%%%%%%%%%
%%%%%%%%%%%%%%%%%%%%%%%%%%%%%%%%%%%%%%%%%%%%%%%%%%%%%

\section{SPIN RESPONSES IN EXCLUSIVE CHANNELS}

The CLAS detector with its large acceptance and good resolution offers
the possibility of studying spin observables for exclusive channels. 
All single pion channels are currently being studied. 
These measurements give information on the isospin dependence of the helicity 
structure in pion production. Since many resonances couple strongly to the 
$N\pi$ channel we also obtain information on the helicity structure 
of resonance excitations. I discuss briefly the $n\pi^+$
 channel where final results have been 
obtained. For details see the talk presented by R. DeVita in the parallel 
session \cite{devita}. 

\begin{figure}[t]
\vspace{100mm} 
\centering{\includegraphics{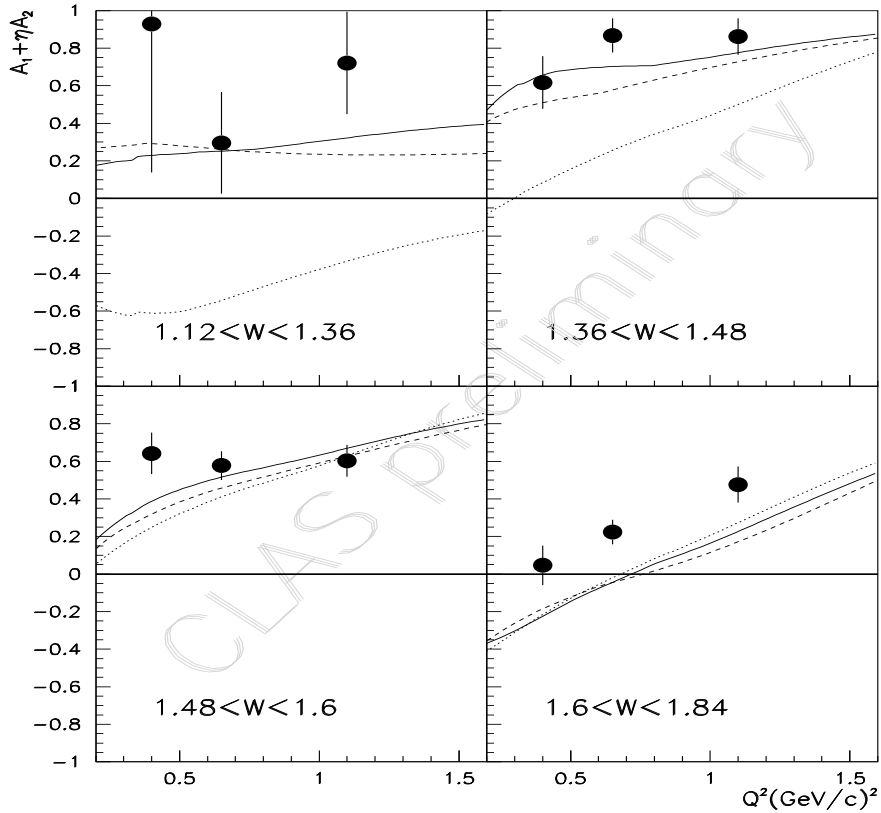}}
\caption{\small Helicity asymmetry $A_1 + \eta A_2$ for the process
 $\vec{p}(\vec{e},e^{\prime}\pi^+)n$. The curves are from parametrizations 
of s-channel resonances and non-resonant Born term contributions to 
single pion production. Solid lines are from the full calculation from 
AO \cite{ao}. Dotted lines are from AO
for resonances only. Dashed lines are the full calculation from 
MAID \cite{maid}.
The various panels are for different ranges in the hadronic mass spectrum.} 
\label{aetpipl}
\end{figure}

Figure {\ref{aetpipl}} shows $A_1+\eta A_2$ for an 
average $Q^2=0.6$GeV$^2$. This channel has a large non-resonant contribution
due to the t-channel pion pole term. Resonance production is, 
therefore, somewhat 
masked. This is particularly evident in the region of the $\Delta(1232)$ where
the measured asymmetry is small. This is in contrast to the previously 
discussed inclusive reaction, and also to the expected asymmetry $A_1 = -0.5$ for a pure
magnetic dipole transition such as the $\Delta(1232)$. 

Above the $\Delta$ mass region, the asymmetry is large and 
positive similar to the inclusive results. Models that include s-channel resonance excitations make predictions
for the $Q^2$-dependence of $A_1 + \eta A_2$  based on the analysis of 
angular distributions of unpolarized differential cross sections \cite{ao,maid}. 
While 
for the $\Delta(1232)$ a rather small asymmetry, with rather weak $Q^2$ 
dependence, is predicted, the higher mass regions should have a large asymmetry
with a strong tendency the towards maximum value of +1. Figure {\ref{aetpipl}} shows
different ranges in W corresponding to different resonance 
regions. The asymmetry in the region of the Roper resonance is nearly  
maximum and dominated by non-resonant contributions. The region near W=1.5GeV
is dominated by the $S_{11}(1535)$ and $D_{13}(1535)$ states, and model parametrizations 
predict large and positive asymmetries rising with $Q^2$. 
This is qualitatively
seen in the data. The largest discrepancy between the data and the model 
calculations is seen in the 3rd resonance region. 
While the trend of the data is reproduced, the resonance 
contributions are significantly underestimated.                   

We also note that $A_1 + \eta A_2$ in the resonance 
region is significantly larger than in the deep inelastic domain.

%%%%%%%%%%%%%%%%%%% %%%%%%%%%%%%%%%%%%%%%%%%%%%%%%%%
%%%%%%%%%%%%%%%%%%%%%%%%%%%%%%%%%%%%%%%%%%%%%%%%%%%%%

\section{Conclusions and Outlook}

First measurements of double polarization asymmetries have been 
carried out at Jefferson Lab in a range of $Q^2$ not covered in high energy
experiments. The results show large contributions from resonance 
excitations with a rapidly changing helicity structure. 
The first moment $\Gamma^p_1(Q^2)$ of the spin structure function 
$g_1(x,Q^2)$ has been extracted. It shows a dramatic change with $Q^2$, 
including a sign change near $Q^2 = 0.3$GeV. Qualitatively, this marks the 
transition from the domain of single parton physics to the domain of
 resonance excitations 
and hadronic degrees of freedom. 

For the first time beam-target asymmetries have been measured 
in exclusive $\pi^+$ electroproduction. Large asymmetries are seen in the
resonance region. With increasing $Q^2$ they tend to rapidly approach the 
limit where helicity 1/2 production dominates the process.

New data have been taken both on hydrogen and deuterium with nearly 10 times
more statistics, and higher target polarizations, and over a larger range 
of energies from 1.6 GeV to 5.75 GeV. These data cover a  
$Q^2$ range from 0.05 to 2.5 GeV$^2$, and a larger part of the deep 
inelastic regime than our previous data. This will allow to reduce the 
systematic uncertainties
related to the extrapolation to $x = 0$. The new data will also give much 
better sensitivity to resonance production in a large number of 
exclusive channels. Finally, at the higher energies, we will be able to 
study single spin asymmetries in various inclusive and exclusive 
reactions.

There is also a program underway in JLab Hall A to measure the GDH 
integral for neutrons down to extremely small $Q^2$ values, 
near the real photons point, and to measure the asymmetry $A_1(x,Q^2)$ 
for the neutron at high x.

\vspace{0.5cm}\noindent
The Southeastern University Research Association (SURA) operates JLab for the 
U.S. Department of Energy under Contract No. DE-AC05-84ER40150.

%\section*{References}

\end{document}